\documentclass[showpacs,preprintnumbers,amssymb]{revtex4}
\usepackage{graphicx}
\usepackage[latin1]{inputenc}
\usepackage{shortvrb}
\usepackage{epsfig}
\usepackage{amssymb}
\usepackage{mathrsfs}
\usepackage{amsbsy}
\usepackage{graphics}
\usepackage{epsf}
\usepackage{amssymb}
\usepackage[usenames]{color}
\usepackage{pstricks}

\usepackage{indentfirst}

\newcommand{\beq}{\begin{equation}}
\newcommand{\eeq}{\vspace{0cm} \end{equation}}
\newcommand{\beqq}{\setlength\arraycolsep{2pt}\begin{eqnarray}}
\newcommand{\eeqq}{\vspace{0cm} \end{eqnarray}}
\newcommand{\beqqq}{\setlength\arraycolsep{2pt}\begin{eqnarray*}}
\newcommand{\eeqqq}{\vspace{0cm} \end{eqnarray*}}
\newcommand{\e}{\textrm{\,e}}
\newcommand{\hsp}{\hspace{1cm}}
\newcommand{\mC}{\mathscr{C}}

\begin{document}

\title{ Hard thermal loops in static external fields }

\author{J. Frenkel$^1$} 
\author{S. H. Pereira$^2$} 
\author{N. Takahashi$^1$}

\vspace{1cm}

\affiliation{\vspace{0.6cm} $^1$Universidade de S\~ao Paulo -- Instituto de F\'isica \\
Rua do Mat\~ao, Travessa R, 187 -- 05508-090 Cidade Universit\'aria, S\~ao Paulo, SP,
Brazil \\ \\  $^2$Universidade de S\~ao Paulo -- Instituto de Astronomia,
  Geof\'\i sica e Ci\^encias Atmosf\'ericas \\
Rua do Mat\~ao, 1226 -- 05508-090 Cidade Universit\'aria, S\~ao Paulo, SP,
Brazil}

\bigskip
\begin{abstract}
We study, in the imaginary-time formalism, the high temperature behavior of $n$-point thermal loops in static
Yang-Mills and gravitational fields. We show that in this regime, any hard
thermal loop gives the same leading contribution as the one obtained by
evaluating the loop at zero external energies and momenta. 
\end{abstract}

\maketitle

\section{Introduction}

In thermal field theory, much attention has been devoted to the study of the high temperature behavior of the Green functions \cite{kala,weld,kaja,hein}, when all their external energies and momenta are much smaller than the temperature $T$. These so-called hard thermal loops \cite{braa,frenkel1} are an important ingredient in a resummation procedure which is necessary to control infrared divergences and give meaningful results in perturbation theory \cite{rebh}. These thermal amplitudes enjoy some simple gauge invariant and symmetry properties, being in general non-local functionals of the external fields. There are two special cases, namely, the static and the long wavelengh limit, when these amplitudes become local functions of the external fields, which are independent of energies as well as of momenta. Nevertheless, these two limits yield in general distinct results for hard thermal self-energy functions \cite{gross,w2,a1,rebhan,a2,arn,a4}. Moreover, these limits also lead to different hard thermal loop effective
actions \cite{tayl, brandt}.

Such a behavior may be more readily understood in the analytically continued imaginary time formalism \cite{kapusta,bellac,das}. In this approach, the bosonic Green functions, for example, are defined at integral values of $k_{l0}/2\pi i T$, where $k_{l0}$ is the energy of the $l$th external particle. Hence, any factors like $\exp(k_{l0}/T)$ can be set equal to unity. This suppresses any factors which could be exponentially increasing after analytic continuation to general values of $k_{l0}$. Then, the integrands of the thermal amplitudes become rational functions when all $k_{l0}$ are complex and various limits can be taken.

Let us consider a typical term which appears in the integrand of any one-loop thermal diagram, namely:
\beq
f(k_0,\vec{k},\vec{Q})={N(k_0 + P)-N(Q)\over k_0+P-Q}\,, \label{eq1}
\eeq
where $N$ is the thermal distribution function for bosons or fermions
\beq
N(z)={1\over \e^{z/T}\mp 1}\,. \label{eq2}
\eeq
$Q=|\vec{Q}|$ is the magnitude of the internal momentum, $P=|\vec{Q}+\vec{k}|$ and $k_0,\, \vec{k}$ are some linear combinations of external energies and momenta. As we have mentioned, before analytic continuation all $k_{l0}$ are integer multiples of $2\pi iT$, so that one may use the relation:
\beq
N(k_0+P)=N(P)\,,\label{eq2A}
\eeq
in which case (\ref{eq1}) can be written in the form:
\beq
\tilde{f}(k_0,\vec{k},\vec{Q})={N(P)-N(Q)\over k_0+P-Q}\,. \label{eq3}
\eeq
However, after analytic continuation, there is another important difference between (\ref{eq1}) and (\ref{eq3}): whereas $f$ is an analytic function at $k_\mu =0$, $\tilde{f}$ is no longer analytic at this point. For instance, in the static limit $k_0=0$, (\ref{eq3}) may be approximated to leading order by $dN(Q)/dQ$. On the other hand, (\ref{eq3}) would vanish in the long wavelength limit $\vec{k}=0$.  Furthermore, it is easy to verify that if we put in the thermal loop, from the start, all $k_{l\mu}=0$, the corresponding term in the integrand would be just:
\beq
f(0,0,Q)={dN(Q)\over dQ}\equiv N'(Q)\,,\label{eq3A}
\eeq
in accordance with the result obtained in the static limit. This agreement occurs because only in the static case, analytic continuation leaves unchanged the original term. Thus, only the local form obtained in the static limit for a general hard thermal loop is equivalent to the result got by setting in the loop all external energies and momenta equal to zero.

The above simple argument can be easily extended to show that the zero energy-momentum limit and the static  limit of any hard thermal loop lead, in fact, to the same result. In section 2, we study the 2-point functions in Yang-Mills and gravity theories, where we try to express the argument in a way which clearly generalizes to higher point functions. Then, in section 3, we further analyse some aspects of the argument in the context of gluon and graviton 3-point functions. We conclude with a brief discussion in section 4.

\section{The 2-point function}

Although the high temperature limit of self-energy functions is well known, we briefly discuss them here in order to present the main points of the argument in a simple form which can be easily generalized. Thus, let us consider the diagram in Fig. 1 where, for our purpose, we need not specify the nature of the particles in the loop. 
\begin{figure}[htb]
\begin{center}
\vspace{0.8cm}
\epsfig{file=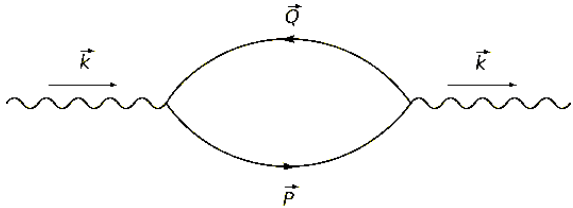, scale=0.4}
\caption{A 2-point self-energy diagram. Wavy lines denote external gluons or gravitons and solid lines indicate internal thermal particles.}\label{fig1}
\end{center}
\end{figure}
Furthermore, since we treat in a unified way thermal loops in either external Yang-Mills or graviton fields, we suppress for simplicity the color and Lorentz indices. Then, the contribution of this diagram may be written in the form:
\beq
\Pi= {\mC_2\over (2\pi)^3}\int d^3 Q\,\, I\,, \label{eq5}
\eeq 
where $\mC_2$ is a Casimir for the internal particles and 
\beqq
I=T\sum_{Q_0}{1\over \vec{Q}^2-Q_0^2}{1\over \vec{P}^2-(Q_0+k_0)^2} t(Q_0)={1\over 2\pi i}\int_C dQ_0 N(Q_0){1\over \vec{Q}^2-Q_0^2}{1\over \vec{P}^2-(Q_0+k_0)^2} t(Q_0)\,.\label{eq6}
\eeqq
Here the sum is over even (odd) integer values of $Q_0/\pi iT$ and $C$ is a contour surrounding all poles of $N$ in an anti-clockwise sense. The numerator $t$ is a tensor and we have indicated, for simplicity, only its dependence on the internal energy $Q_0$. In fact, in the Yang-Mills theory $t$ is a quadratic function of the energies and momenta, whereas in gravity $t$ becomes a function of fourth degree in the energies and momenta. Evaluating (\ref{eq6}) in terms of the poles outside $C$, and writing $Q=|\vec{Q}|$, $P=|\vec{P}|$, we get:
\beqq
I&=&-{1\over 4PQ}\Bigg\{N(Q)t(Q)\Big[{1\over Q-P+k_0}-{1\over Q+P+k_0}\Big]+(Q,P\to -Q,-P)\nonumber\\
&&+N(P)t(P-k_0)\Big[{1\over P-Q-k_0}-{1\over P+Q-k_0}\Big]+(Q,P\to -Q,-P)\Bigg\}\,,\label{eq7}
\eeqq
where we used the relation $N(P-k_0)=N(P)$ etc, since in the imaginary time formalism $k_0/2\pi iT$ is an integer.

We now consider the leading high temperature contribution in the static case ($k_0=0$), which comes from the region $|\vec{k}|\ll P,Q\sim T$. To this end, we can make appropriate expansions like:
\beq
t(P)= t(Q)+(P-Q)t'(Q)+\cdots \label{eq8}
\eeq
and similar ones for $N(P)$. Then, it is easy to check that to leading order, (\ref{eq7}) reduces to:
\beq
I_S= -{1\over 4 Q^2}\Bigg\{[N'(Q)-{N(Q)\over Q}][t(Q)+t(-Q)]+N(Q)[t'(Q)-t'(-Q)]\Bigg\}\,,\label{eq9}
\eeq
where we used the relation
\beq
 N(Q)+N(-Q)\pm 1=0\label{eq10b}
\eeq
and omitted a $T$-independent term. At this point we note that in the integral (\ref{eq5}), only those components with an even number of $Q_i$, and hence an even number of $Q_0$, do actually contribute. Thus $t(Q_0)$ becomes effectively an even function of $Q_0$ while $t'(Q_0)$ becomes an odd function, so that one may further simplify (\ref{eq9}) as:
\beq
I_S=-{1\over 2Q^2}\bigg[N'(Q)t(Q)+N(Q)t'(Q)-{N(Q)\over Q}t(Q)\bigg]\,.\label{eq10}
\eeq

Now, let us compare this result with the one obtained by setting, from the start, $k_\mu=0$ in the diagram in Fig. 1. This gives (compare with (\ref{eq6})):
\beq
I_0={1\over 2\pi i}\int_C dQ_0 N(Q_0){1\over (Q_0^2-Q^2)^2}t(Q_0)\,.\label{eq11}
\eeq

We can evaluate the $Q_0$ integral in (\ref{eq11}) by residues, in terms of the double poles outside $C$. Using the properties mentioned above and omitting a $T$-independent term, one readily gets the result:
\beq
I_0=-2{d\over dQ_0}\Bigg[{N(Q_0)t(Q_0)\over (Q_0+Q)^2}\bigg]_{Q_0=Q}\,.\label{eq12}
\eeq
This is actually equal to the result given in (\ref{eq10}), as expected from the argument given in the previous section.

Finally, after performing the $d^3Q$ integration in (\ref{eq5}), one obtains the well known leading $T^2$ contribution for the static gluon 2-point function, while the static graviton 2-point function gives a leading contribution of order $T^4$.

\section{The 3-point function}

Consider the triangle diagram shown in Fig. 2. Its contribution in the imaginary time formalism may be written in the form:
\beq
\Gamma={\mC_3\over (2\pi)^4 i}\int d^3Q\int_C dQ_0 N(Q_0){1\over \vec{Q}^2 - Q_0^2}{1\over \vec{R}^2 - (Q_0+k_{10})^2}{1\over \vec{P}^2 - (Q_0-k_{30})^2}t(Q_0)\,. \label{eq13}
\eeq
Here the Casimir $\mC_3$ gives the number of internal degrees of freedom. 
\begin{figure}[htb]
\begin{center}
\vspace{0.8cm}
\epsfig{file=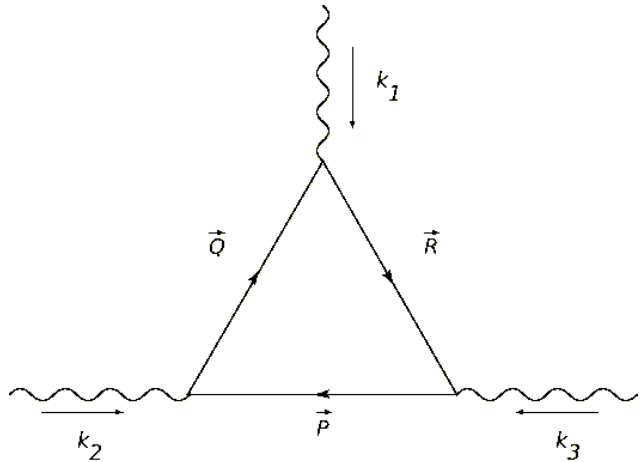, scale=0.45}
\caption{One loop 3-point vertex diagram. Colour and Lorentz indices are suppressed.}\label{fig2}
\end{center}
\end{figure}
The tensor $t$, whose colour and Lorentz indices have been suppressed for conciseness, is a function of the energies and momenta of degree three or six in the Yang-Mills or gravity theory, respectively. Evaluating the $Q_0$ integral by residues and using $N(P+k_{30})=N(P)$ etc, we obtain
\beq
\Gamma = -{1\over (2\pi)^3}\int d^3Q \bigg[I+I'-(I_1+I'_1 +I_2+I'_2+I_3+I'_3)\bigg]\,,\label{eq14}
\eeq
where
\beq
I ={1\over 8PQR}\Bigg[ {t(Q)N(Q)\over D_3 D_1} + {t(P+k_{30})N(P)\over D_2 D_3}+{t(R-k_{10})N(R)\over D_1 D_2}\Bigg]\,,\label{eq15}
\eeq
\beq
I_1 ={1\over 8PQR}\Bigg[ {t(Q)N(Q)\over D_3 E_1} + {t(P+k_{30})N(P)\over D_3 F_2}+{t(-R-k_{10})N(-R)\over E_1 F_2}\Bigg]\,.\label{eq16}
\eeq
Here we have used the notation
\beq
D_1=k_{10}+Q-R\,, \hsp E_1=k_{10}+Q+R\,,\hsp F_2=k_{20}-R-P\label{eq17}
\eeq
and cyclic permutations ($k_1\to k_2\to k_3$ and $Q\to P\to R$), so that
\beq
D_1 + D_2 + D_3=0\,.\label{eq18}
\eeq
$I_2$ and $I_3$ are obtained by cyclic permutations on (\ref{eq16}) and:
\beq
I'(P,Q,R)=I(-P, -Q, -R)\,, \hsp I'_1(P,Q,R)=I_1(-P, -Q, -R)\,.\label{eq19}
\eeq
In the static limit $k_{i0}=0$, the leading contributions at high $T$ come from the region $|\vec{k}_i|\ll P,\,Q,\,R\sim T$. One can thus make appropriate expansions like
\beq
t(P)=t(Q)+(P-Q)t'(Q)+{1\over 2}(P-Q)^2t''(Q)+ \cdots \label{eq20}
\eeq

Let us first consider the contributions to (\ref{eq14}) associated with the $t(Q)$ term in the above expansions. Then, using similar expansions for $N(P)$, etc as well as the relation (\ref{eq18}), one finds that the corresponding terms in (\ref{eq15}), (\ref{eq16}) and (\ref{eq19}) give:
\beq
I_a=I'_a = -{1\over 16}{N''(Q)\over Q^3} t(Q)\,, \label{eq21}
\eeq
\beq
I_{1a}=I'_{1a} = {1\over 16}{N(Q)-QN'(Q)\over Q^5} t(Q)\,. \label{eq22}
\eeq
Inserting (\ref{eq21}), (\ref{eq22}), etc into (\ref{eq14}), we get the contribution
\beq
\Gamma_a={\mC_3\over 8} \int {d^3Q\over (2\pi)^3}{1\over Q^5} \bigg[Q^2 N''(Q)-3QN'(Q) + 3N(Q)\bigg] t(Q)\,.\label{eq23}
\eeq

We must now include the contributions to (\ref{eq14}) associated with the terms $t'(Q)$ and $t''(Q)$ in expansions like (\ref{eq20}). The calculation is straightforward and the corresponding expressions are:
\beq
\Gamma_b={\mC_3\over 8} \int {d^3Q\over (2\pi)^3}{1\over Q^4} \bigg[2QN'(Q) - 3N(Q)\bigg] t'(Q)\,,\label{eq24}
\eeq
\beq
\Gamma_c={\mC_3\over 8} \int {d^3Q\over (2\pi)^3}{1\over Q^3} N(Q)t''(Q)\,.\label{eq25}
\eeq
Thus, in the static case, the three terms (\ref{eq23}), (\ref{eq24}) and (\ref{eq25}) contribute to (\ref{eq14}) as
\beq
\Gamma_S={\mC_3\over 8} \int {d^3Q\over (2\pi)^3}{1\over Q^5} \Bigg\{Q^2\bigg[ N''(Q)t(Q)+2N'(Q)t'(Q) + N(Q)t''(Q)\bigg]-3Q\bigg[N'(Q)t(Q)+N(Q)t'(Q)\bigg]+3N(Q)t(Q)\Bigg\}\,.\label{eq26}
\eeq

Let us next compare this result with the one obtained from the diagram in Fig. 2,  when all external energies and momenta are vanishing. We then get:
\beq
\Gamma_0={\mC_3\over (2\pi)^4i}\int d^3Q \int_C dQ_0 N(Q_0){1\over (Q^2-Q_0^2)^3}t(Q_0)\,. \label{eq27}
\eeq
One can evaluate the $Q_0$ integral by residues, in terms of the triple poles outside the contour $C$. Then, using the relation (\ref{eq10b}) and the properties mentioned afterward, we can write $\Gamma_0$ in the simple form:
\beq
\Gamma_0=\mC_3 \int {d^3 Q\over (2\pi)^3}\, {d^2\over dQ_0^2}\Bigg[{N(Q_0)t(Q_0)\over (Q_0 + Q)^3}\Bigg]_{Q_0=Q}\,. \label{eq28}
\eeq
This result is in fact equivalent to the one in (\ref{eq26}), as expected from the general argument given in section 1. Clearly, this feature will also hold for higher-point functions.

\section{Discussion}

We examined, in the analytically continued imaginary-time formalism, the behavior of hard thermal loops in static external Yang-Mills and gravitational fields. Amplitudes calculated in this formalism naturally give rise to retarded (advanced) hard thermal functions \cite{evans}, whose imaginary parts vanish in the static limit. We argued that for any hard thermal loop, the leading contributions in the static limit are the same as those obtained at zero external energy-momentum. This is consistent with the behavior of thermal self-energy loops noticed in \cite{gross,w2,a1,rebhan,a2,arn,a4}. This result may be useful to simplify the calculation of static limits in thermal field theories, which are relevant to study some physical properties of systems in thermal equilibrium, like plasma frequencies and screening lengths. 

Although the above relationship holds both in the Yang-Mills as well as in the gravity theory, there is an important difference between them. One can show \cite{tayl} that in the Yang-Mills case, all higher order point functions vanish to leading order in the static limit. This fact can also be simply understood from our previous argument since, after setting in the thermal loop all external energies and momenta equal to zero, one can see by power counting that higher point functions can no longer yield quadratic $T^2$ contributions. This implies, in particular, that (\ref{eq28}) should vanish in Yang-Mills theory, a fact that also follows from Bose symmetry. Indeed, there appears in this case an antisymmetric colour factor which requires another factor with an antisymmetric dependence on external momenta and energy. But such a factor will necessarily vanish in the zero energy-momenta limit.

On the other hand, in gravity, such an antisymmetric dependence cannot be present and then, since in (\ref{eq28}) $t$ is a function of sixth degree in $Q$, it will yield a leading contribution of order $T^4$. The fact that there are, for all $n$-point functions, static $T^4$ terms in gravity is of course connected with the quartic ultraviolet divergence of the zero-temperature loops. These static thermal functions are related by Ward
identities \cite{frenkel2}, which are associated with the invariance of this system under time-independent
coordinate transformations.

\section*{Acknowledgements}

JF is grateful to J. C. Taylor for a very helpful correspondence. He would also like to thank F. T. Brandt for useful discussions. SHP is supported by CNPq, proc. num. 150920/2007-5. This work was supported in part by CNPq and FAPESP, Brazil.

\end{document}